# Interplay of pH and Binding of Multivalent Metal Ions: Charge Inversion and Reentrant Condensation in Protein Solutions

Felix Roosen-Runge,[1, a)] Benjamin S. Heck,[1] Fajun Zhang,[1] Oliver Kohlbacher,[2] and Frank Schreiber[1, b)]
[1)]*Institut für Angewandte Physik, Universität Tübingen, Auf der Morgenstelle 10, 72076 Tübingen, Germany*
[2)]*Zentrum für Bioinformatik, Zentrum für quantitative Biologie und Fachbereich Informatik, Universität Tübingen, Sand 14, 72076 Tübingen, Germany*

(Dated: 17 April 2013)

Tuning of protein surface charge is a fundamental mechanism in biological systems. Protein charge is regulated in a physiological context by pH and interaction with counterions. We report on charge inversion and the related reentrant condensation in solutions of globular proteins with different multivalent metal cations. In particular, we focus on the changes in phase behavior and charge regulation due to pH effects caused by hydrolysis of metal ions. For several proteins and metal salts, charge inversion as measured by electrophoretic light scattering is found to be a universal phenomenon, the extent of which is dependent on the specific protein-salt combination. Reentrant phase diagrams show a much narrower phase-separated regime for acidic salts such as $AlCl_3$ and $FeCl_3$ compared to neutral salts such as $YCl_3$ or $LaCl_3$. The differences between acidic and neutral salts can be explained by the interplay of pH effects and binding of the multivalent counterions. The experimental findings are reproduced with good agreement by an analytical model for protein charging taking into account ion condensation, metal ion hydrolysis and interaction with charged amino acid side chains on the protein surface. Finally, the relationship of charge inversion and reentrant condensation is discussed, suggesting that pH variation in combination with multivalent cations provides control over both attractive and repulsive interactions between proteins.

## I. INTRODUCTION

Tuning of protein surface charge is a fundamental mechanism ensuring protein stability and function in aqueous solutions[1–3]. The protein surface charge stabilizes protein solutions[4,5] and cluster phases[6,7]. Non-homogeneous surface charge distributions represent anisotropic interaction patches which crucially affect phase behavior of solutions[8,9] as well as pathways for aggregation and crystallization[10,11].

In electrolyte solutions, control on the protein charge is realized mainly by two processes: (de)protonation of functional surface groups and counterion condensation on the protein surface. Protonation and deprotonation, as investigated in titration experiments[12,13], modulates the charge state of basic and acidic amino acid side chains (Lys, Arg, Glu, Asp, His) as well of the protein termini (carboxy and amino terminus). As in every proteolytic equilibrium, the charge state of each side chain – and thus the overall protein charge – depends on the pH of the surrounding medium.

Upon adding salt ions, charge regulation becomes more complex due to the ion-protein interaction. Many phenomena such as ion condensation[14] and ion binding[15] have been found to play a role in protein-ion interaction and charge regulation[16]. The discovery of the Hofmeister series[17] inspired an enduring effort to understand the protein-salt interaction, focussing on surface hydrophobicity[18], ion hydration[19] and ion polarizability[20]. In fact, ion specific effects on the surface charge of proteins in electrolyte solutions are still a challenge for theory[21]. The association of salt counterions with inversely charged side chains at the protein surface seems to be the dominant effect and is suggested as the basic model for ion-protein interaction[22]. Ion association has been identified to occur mainly at hydrophilic sites surrounded by hydrophobic surface areas[23]. Amino acids with carboxylate, thiol, thioether, and imidazole groups can coordinatively bind transition metal ions[24,25]. Taking into account also the charge of the molecule, binding models have provided the basis for the understanding of interactions of proteins with ions and ligands[15,26–28].

Importantly, also non-local effects, i.e. not only dependent on one single functional group, affect ion association. First, since the protein/water interface exhibits a change of dielectric constant, also non-localized adsorption of polarizable, i.e. large, ions could occur at non-polar, hydrophobic areas of the protein surface[22,29]. Second, ion-ion correlations and finite size effects close to surfaces with high charge density have been shown to cause overcharging of the surface, attraction between like-charged surfaces, and charge inversion of colloidal particles[30–35].

A complete model for the protein charge in any system has to include both local and non-local contributions, although many studies focus on only one approach[36,37].

The driving force of ion-induced surface charge is the change in free energy upon ion condensation on the surface, which can be described qualitatively – and to a reasonable degree also quantitatively – with an effective short-ranged interaction between protein and ions, even for long-ranged and non-localized interactions[37].

In this article, we present findings on the charge inver-

---

[a)] felix.roosen-runge@uni-tuebingen.de
[b)] frank.schreiber@uni-tuebingen.de

sion of globular proteins, in particular bovine and human serum albumin (BSA, HSA), ovalbumin (OVA) and β-lactoglobulin (BLG), in the presence of multivalent metal cations. We focus on the effects of the trivalent metal ions $Al^{3+}$, $Fe^{3+}$ and $Y^{3+}$ and, in particular, the effects of pH variations arising from the metal hydrolysis. The charge inversion is reflected in a reentrant condensation of the protein solution[38–40], i.e. the protein solution is homogeneous and stable for low and high metal ion concentrations, while for intermediate metal salt concentrations a phase-separated state is found, corresponding to either a liquid-liquid phase separation or formation of amorphous clusters and crystals[41,42]. These phenomena do not occur in solutions of monovalent cations and cannot be understood by conventional salting-in or salting-out effects[43–45].

After an outline of the experimental methods, we present experimental results on reentrant condensation in protein solutions with trivalent salts and the related inversion of the protein charge. We introduce a model for metal ion association and hydrolysis, which conceptually explains the observed phenomena. The theoretical methods are summarized in more detail in the Appendix.

## II. EXPERIMENTAL METHODS

### A. Sample preparation and optical characterization

All proteins were purchased from Sigma-Aldrich (BSA: 99 % purity, A3059; HSA: 97–99 %, A9511; BLG: 90 %, L3908; OVA: 98 %, A5503) and used as received. $FeCl_3$, $AlCl_3$ and $YCl_3$ were purchased from Sigma Aldrich at high purity grade (>99 %). Deionized water was used to prepare all solutions.

Stock solutions for proteins (200 mg/ml) and salts (200 mM) were prepared and then diluted in order to obtain the desired concentrations. The protein stock solution was filtered through 0.22 $\mu$m membrane filters (Millipore). No buffer was used in order to study solely the effect of the salt in the protein-protein interactions. pH of freshly prepared protein-salt solutions was measured using the pH meter Mettler Toledo InLab 413 SG/2n IP67, which was calibrated before measurement to standard buffer solutions. The phase behavior of protein solutions was determined using the optical transmission of the solutions[39] or by visual inspection. All preparations and measurements were performed at room temperature.

In order to ensure structural stability of the proteins at the given metal ion concentration and pH, the secondary structure content was monitored using Fourier Transform Infrared spectroscopy. No significant changes have been found in the amide I ($\approx 1650\,\mathrm{cm}^{-1}$) and amide II bands ($\approx 1450\,\mathrm{cm}^{-1}$) at different metal ion concentrations, implying that the secondary structure of the protein remains invariant for the range of salt concentration studied.

### B. Electrophoretic Light Scattering

All measurements of electrophoretic mobility were performed with a Zetasizer Nano (Malvern Instruments GmbH, Germany), using electrophoretic light scattering (ELS) via phase analysis light scattering (PALS). From the electrophoretic mobility, the so-called zeta potential $\zeta$ can be calculated, which is the potential at the outer border of the hydrodynamic stagnant layer around the protein. In general, $\zeta$ reflects the total charge within the stagnant layer, but the analytical relation is not trivial in the case of soft, rough and non-spherical objects[46]. The next paragraphs outline the calculation of the protein charge from ELS measurements. Note that this approach is based on several assumptions, most importantly a mean-field theoretical approach for the ion distribution and an effective sphere as representation of the protein. Since both assumptions are not fully fulfilled in our system, the resulting charges should be considered as effective charges[47] which are not absolutely precise, but suitable for relative comparison and further discussion.

ELS measures the particle mobility $\mu_E$ in an external electrical field which is related to the zeta potential for a spherical particle with radius $a$:[46]

$$\mu_E = \frac{2\varepsilon}{3\eta} f(\kappa a) \zeta \; . \quad (1)$$

$\varepsilon$ is the total dielectric permittivity and $\eta$ the viscosity of the medium. The inverse Debye screening length $\kappa$ and the Bjerrum length $\lambda_B$ represent the relevant length scales for the electrostatic interaction:

$$\kappa^2 = 4\pi \lambda_B \sum_i n_i Z_i^2 \quad , \quad \lambda_B = e^2/(\varepsilon k_B T) \; . \quad (2)$$

Here, all present ionic species $i$ are accounted for, with the respective valency $Z_i$ and number concentration $n_i$. $e$ denotes the elementary charge and $k_B T$ the Boltzmann constant multiplied with the temperature.

The Henry function $f(\kappa a)$ in Eq. (1) interpolates between the Helmholtz-Smoluchowski limit ($\kappa a \gg 1$: $f = 1.5$) and the Hückel-Onsager limit ($\kappa a < 1$: $f = 1$). $f(\kappa a)$ can be approximated by Oshima's relation[46,48]:

$$f(\kappa a) = 1 + \frac{1}{2} \left[ 1 + \left( \frac{2.5}{\kappa a [1 + 2\exp(-\kappa a)]} \right) \right]^{-3} \; . \quad (3)$$

The reduced zeta potential $\tilde{\zeta} = \frac{e\zeta}{2k_B T}$ is related to the surface charge density $\sigma$:[49]

$$\frac{\sigma e}{k_B T} = 2\epsilon\kappa \left[ \sinh^2\left(\frac{\tilde{\zeta}}{2}\right) + \frac{2}{\kappa a}\tanh^2\left(\frac{\tilde{\zeta}}{4}\right) \right.$$
$$\left. + \frac{8}{(\kappa a)^2} \ln\left(\cosh\left(\frac{\tilde{\zeta}}{4}\right)\right) \right]^{1/2} \; . \quad (4)$$

Based on $\sigma$ we calculate the total charge of the protein-salt complex from

$$Q = 4\pi a^2 \sigma \quad (5)$$

For the different proteins we estimated the following effective radii[45]: $a_{BSA} = a_{HSA} = 3.3\,\text{nm}$, $a_{BLG} = 2.7\,\text{nm}$, $a_{OVA} = 3.6\,\text{nm}$. For BLG and OVA, we considered dimers.

## III. EXPERIMENTAL RESULTS

### A. Reentrant Condensation

1 summarizes the phase diagrams determined by optical transmission and visual inspection showing reentrant condensation behavior for BSA with three different metal ions. In earlier studies, reentrant condensation has been observed as a universal phase behavior for several globular proteins with negative native charge (e.g. BSA, HSA, OVA, BLG) in the presence of multivalent metal cations (e.g. $Al^{3+}$, $Fe^{3+}$, $Y^{3+}$, $La^{3+}$)[38,39]. At low salt concentrations, solutions are clear; above a critical salt concentration $c^*$ protein solutions enter a phase-separated state. Further increase of the salt concentration results in a clear solution again.

In the context of this study, it is important to emphasize that there are significant differences concerning the phase-separated regime (1). Quantitatively, the concentration range for the phase-separated regime is clearly narrower for $Al^{3+}$ and $Fe^{3+}$ than for $Y^{3+}$. Qualitatively, the nature of the phase separation differs: while for $AlCl_3$, gels and aggregates are formed, only amorphous aggregates are found in the case of $FeCl_3$. For $YCl_3$, clustering, amorphous aggregates and liquid-liquid phase separation are observed[41].

### B. Inversion of Surface Charge

2 presents the inversion of surface charge for several typical sets of protein-metal ion solutions. The surface charge is calculated by Eq. (4, 5) from ELS measurements. For all studied combinations of proteins with acidic pI (here: BSA, HSA, BLG, OVA) and multivalent metal salt (here: $YCl_3$, $AlCl_3$, $FeCl_3$), a clear inversion of the surface charge is observed upon increasing the metal ion concentration. In 2(top), we observe a clear dependence of the point of zero charge $c_0$ on the protein concentration. This effect is expected, since the higher number density of protein requires a higher concentration of metal ions to compensate for the native protein surface charge.

2(bottom) illustrates the deviations of the quantitative extent of charge inversion with respect to the combination of protein and metal salt. First, the maximum surface charge after charge inversion depends on the protein. This finding is expected, since the number of association sites for metal ions will differ with protein size and structure. While BSA and HSA have comparable size and show comparable absolute numbers, the OVA dimer is significantly larger and apparently obtains higher surface charges after charge inversion. Second, the nature of the metal salt has an effect. Different association behavior is possible, and pH effects due to metal salt hydrolysis are expected to contribute to the charge inversion. The strength of the pH effect is reflected in the large shift of $c_0$ between rather neutral salts (YCl$_3$) and acidic salts (AlCl$_3$, FeCl$_3$).

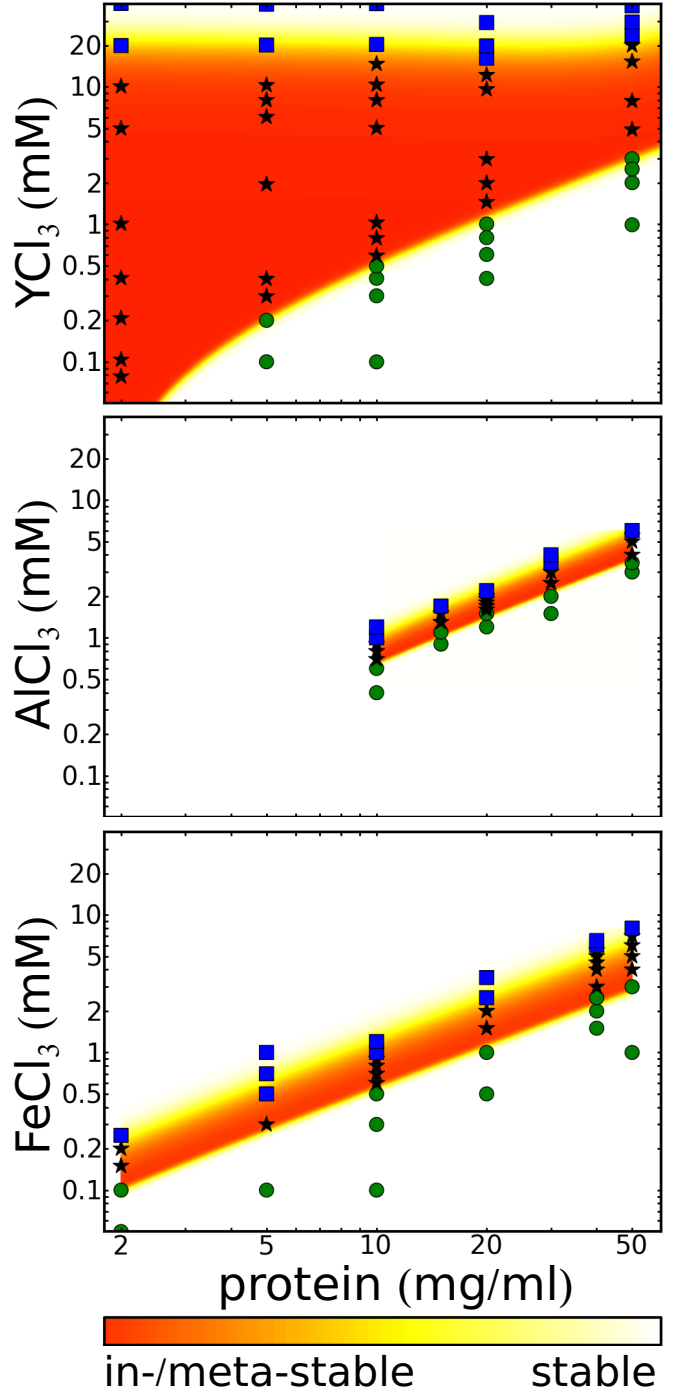

FIG. 1. Reentrant phase behavior for BSA and different multivalent metal ions ($Al^{3+}$, $Fe^{3+}$, $Y^{3+}$). The protein solution is stable for low (green circles) and high (blue squares) metal ion concentrations (white color-coding). For intermediate metal ion concentrations, an unstable or metastable protein solution exists (black stars and red color-coding). Clear differences can be observed between the different metal ions. In particular, the condensed regime occurs on a much narrower metal concentration range for $Al^{3+}$ and $Fe^{3+}$ than for $Y^{3+}$, which is an indication of relevant pH effects due to metal ion hydrolysis. Part of the data are taken from earlier studies.[38,39]

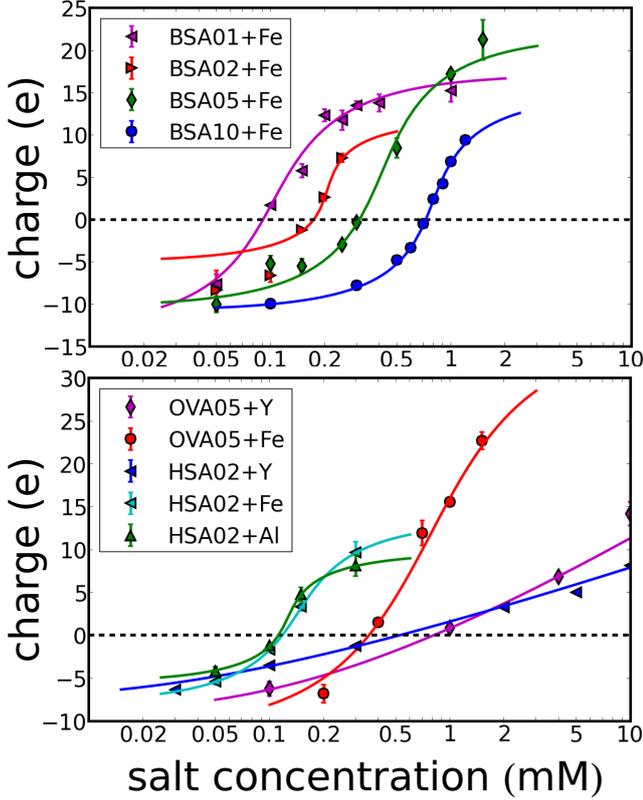

FIG. 2. Inversion of protein surface charge induced by multivalent metal ions as calculated from electrophoretic measurements. All solid lines correspond to Langmuir isotherms with an additional cooperativity parameter (Eq. (6)). Top: Charge inversion for protein concentration series (1, 2, 5, 10 mg/ml BSA) with FeCl$_3$: With increasing protein concentrations, the point of zero charge shifts to higher salt concentrations. The higher concentration of proteins requires a higher amount of metal ions to compensate the native surface charge. Bottom: Comparison between different salts for OVA (5 mg/ml) and HSA (2 mg/ml) with FeCl$_3$, AlCl$_3$ and YCl$_3$: The charge inversion is universally present in all measured samples of BSA, HSA, BLG and OVA in presence of FeCl$_3$, AlCl$_3$ and YCl$_3$. In general, the charge inversion occurs at lower salt concentration for acidic salts like FeCl$_3$ and AlCl$_3$ compared to YCl$_3$, indicating a relevant pH effect due to metal salt hydrolysis.

### C. Semiempirical Fit of Surface Charge

The solid lines in 2 represent fits with a Langmuir-like equation for the protein charge upon variation of salt concentration. The starting point for fitting these charge profiles is the well-known Langmuir isotherm for the association of a ligand at $N$ independent sites with association constant $K$. In our case, however, two mechanisms are expected to change this profile. First, each counterion on the surface will change the protein charge and thus suppress further ion association. Second, the hydrolysis of the metal salts lowers the pH in the experimental concentration range, causing an additional positive charge due to protonation of acidic residues.

In order to semiempirically allow for these two mechanisms, i.e. competitive binding and pH effects, the association constant $K$ is rewritten as a product of a intrinsic association constant $K'$ and an exponential, representing the first-order correction of the free energy with respect to $Q$: $K(Q) = K' \cdot \exp(pQ)$. The cooperativity parameter $p$ corresponds to the free energy per protein charge. Introducing this modification into the Langmuir isotherm, the fit function reads:

$$Q = Q_0 + \frac{ZNc}{c + K' \cdot \exp(pQ)} \ . \qquad (6)$$

$Z$ is the valency of the metal ions, $c$ is the metal ion concentration, $Q_0$ is the protein charge without metal ions and $N$ is the number of accessible association sites. The exponential factor renders the equation implicit in the desired variable $Q$; the solution is obtained from a numerical bisection with respect to $Q$ using the routine supplied by the python package *scipy.optimize*[50,51] (version 0.7.0).

Using Eq. 6, the charge profiles from ELS can be fitted with reasonable accuracy (2). Except from some unphysical fits due to insufficient range of metal concentration, the fit parameters show a consistent behavior: the protein charge $Q_0$ lies between -5 and -12. The number of association sites $N$ varies between 4 and 11. Interestingly, the parameter $p$ is mainly negative. The binding of counterions to the protein, however, is a competitive binding due to the surface charge variation, implying a positive $p$. The negative value of $p$ thus implies a non-negligible pH effect on the charge inversion.

### D. pH of protein solutions with multivalent salts

3 summarizes some results on the pH values of protein solutions with acidic multivalent salts. As expected, the pH decreases with increased salt concentration down to pH 4. At higher protein concentration, the pH decreases more slowly, presumably due to self-buffering of the protein and binding of multivalent salt. Along with the measured pH values, the point of zero charge $c_0$ (red circles), as taken from the fits, and the phase boundaries of the phase-separated regime, $c^*$ and $c^{**}$ (yellow stars), are shown. Importantly, both charge inversion and reentrant condensation take place at pH values above the isoelectric point pI $\approx$ 4.7 of BSA[52]. The charge inversion thus cannot be driven solely by pH variation of the solution[39]; binding of multivalent counterions has also a non-negligible effect.

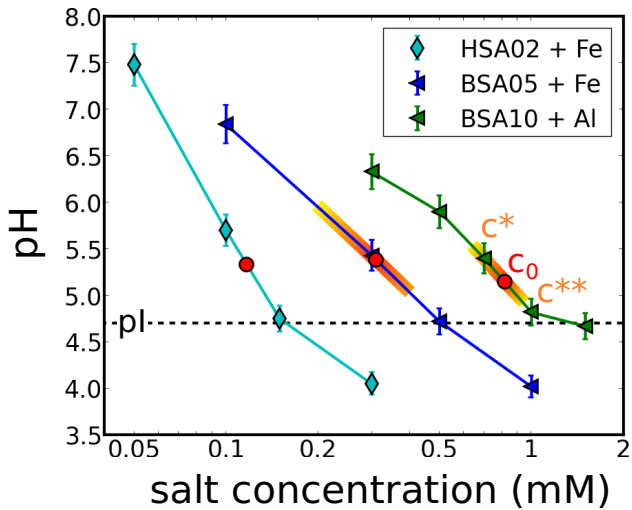

FIG. 3. pH of protein solutions (HSA 2 mg/ml, BSA 5 mg/ml and 10 mg/ml) with FeCl$_3$ and AlCl$_3$. Both charge inversion, indicated by the estimated point of zero charge $c_0$ (red circle), and reentrant condensation, indicated by the critical salt concentrations $c^*$ and $c^{**}$ (colored area), occur at pH values above the isoelectric point pI (black dashed horizontal line). Thus, binding of counterions has a non-negligible effect on the charge regulation.

## IV. ANALYTICAL MODEL FOR THE INTERPLAY OF ION ASSOCIATION AND pH EFFECTS

The experimental results in the previous section indicate that both pH and binding of multivalent ions play an important role for the protein surface charge. Inspired by these experimental findings, we present an analytical model which incorporates several association reactions (for a schematic representation, see 4). (De)Protonation and association of metal counterions take place at the same functional groups at the protein surface and are thus in competition. Due to the geometrical compactness of the protein, the functional groups are not independent, but coupled to each other since the protein total charge contributes non-negligibly to the association free energy. Furthermore, the solution chemistry adds another level of complexity due to the hydrolysis of metal salts and the related changes of the pH in the protein solution. In essence, there is a interplay of protein surface charge, solution pH and salt concentration.

Using a model of a spherical particle in a solution, we aim to extract some essential features by solving a system of association reactions while simultaneously cor-recting for charge and pH effects on the association equilibria. Note that the model should not be regarded as a detailed quantitative description, which would obviously fail, amongst other factors, for the reasons of missing anisotropy effects from non-spherical particle geometry and mean-field approximations for the ion distributions. However, general trends of ion association, interwoven with pH effects, should be accessible from this model which we believe to capture the essential points.

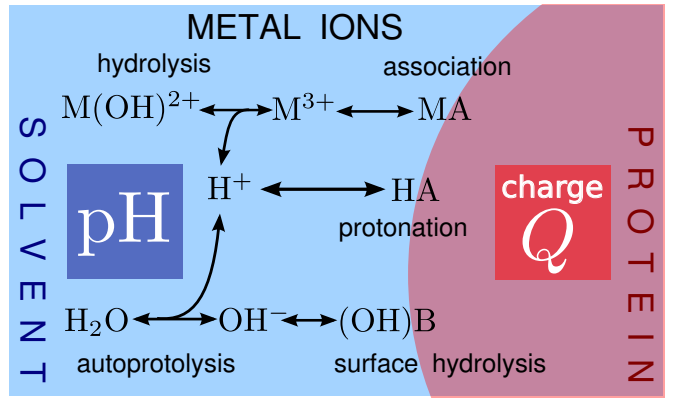

FIG. 4. Schematic representation of the reactions incorporated in the suggested model. The surface equilibria of metal association, protonation, and hydrolysis link the total particle charge to the solution pH. Metal ions with pronounced pH effect due to hydrolysis thus have a non-trivial effect on the surface charge.

Concerning the surface association, the calculation scheme is inspired by adsorption models for metal complexes at surfaces[53,54] as well as ideas from well-known titration calculations for proteins[12,13,55] and models for ion binding[15,26–28,56]. The hydrolysis of the trivalent metal salts used here is well-characterized[57–60]. To our knowledge, so far neither simulation nor theoretical models have addressed the coupling of metal salt hydrolysis and surface association as present in unbuffered solutions of proteins with salts.

### A. Association reactions

When considering a protein in aqueous solution, first and importantly, the charge regulation via acidic and basic residues has to be accounted for. The basic residues are denoted as B$^+$. Not all acidic residues are expected to act as metal association site, e.g. for steric reasons. D$^-$ represents acidic residues acting only as protonation site while A$^-$ denotes a site with both protonation and metal association.

$$\begin{aligned} \text{HA} &\stackrel{K_a}{\rightleftharpoons} \text{A}^- + \text{H}^+ \\ \text{HD} &\stackrel{K_a}{\rightleftharpoons} \text{D}^- + \text{H}^+ \\ \text{BH}^+ &\stackrel{K_b}{\rightleftharpoons} \text{B} + \text{H}^+ \end{aligned} \qquad (7)$$

This reaction has to be complemented by the autoprotolysis of water

$$H_2O \stackrel{K_w}{\rightleftharpoons} OH^- + H^+ \quad (8)$$

Adding multivalent metal ions to the solution adds further association reactions. The hydrolysis of metal ions varies the solution pH, which in turn affects the charge regulation of the protein.

$$M^{3+} + H_2O \stackrel{K_0}{\rightleftharpoons} M(OH)^{2+} + H^+ \quad (9)$$
$$M(OH)^{2+} + H_2O \stackrel{K_1}{\rightleftharpoons} M(OH)_2^+ + H^+$$
$$M(OH)_2^+ + H_2O \stackrel{K_2}{\rightleftharpoons} M(OH)_3 + H^+$$

We use a simplified notation where $M^{3+}$ represents possible metal aquo complexes with the given charge, for example the octahedral hexaaquaaluminium complex ion $Al(H_2O)_6^{3+}$, for other valencies analogously[61]. The limited solubility of $M(OH)_3$ complexes will be discussed in the next subsection. We have neglected further species like $M(OH)_4^-$ since these will play a role in basic solutions only[61].

Finally, binding of metal ions to acidic sites at the protein surface (each specimen, respectively) modulates the surface charge:

$$M^{3+} + A^- \stackrel{K_{m0}}{\rightleftharpoons} MA^{2+} \quad (10)$$
$$M(OH)^{2+} + A^- \stackrel{K_{m1}}{\rightleftharpoons} M(OH)A^+$$
$$M(OH)_2^+ + A^- \stackrel{K_{m2}}{\rightleftharpoons} M(OH)_2A$$
$$M(OH)_3 + A^- \stackrel{K_{m3}}{\rightleftharpoons} M(OH)_3A^-$$

We do not incorporate other association reactions of monovalent ions (like $Na^+$ and $Cl^-$) with both protein and metal salt, since these reactions would unnecessarily complicate the model. For simplicity we assume complete solvation of salt and protein. Note that monovalent ions are implicitly accounted for by the ionic strength which screens the electrostatic contributions (see subsections on free energy contributions and on phase behavior from a DLVO picture). Although assumptions on the ionic strength might quantitatively change the model parameters, the qualitative behavior from the binding of trivalent salt ions and their coupling with the pH is expected to be similar.

## B. Equilibrium constants and solubility

The association reactions mentioned in the previous subsection incorporate several equilibrium constants $K_i$ which have to be chosen reasonably. While the equilibrium constants for the aqueous chemistry can be obtained from literature values, an estimation of equilibrium constants for association reactions on the protein surface is more involved. In principle, the association constant for metal ions and protonation at the protein surface is locally varying with a rather broad distribution. The local shifts arise from inhomogeneities in the electrostatic surface potential and solvation energy due to the local surface morphology as well as possible multi-dentate binding configurations. Since we are aiming for a qualitative understanding of the pH effects on the charge inversion in a simplified picture, we use a single value to represent a hypothetical homogeneous association process of each species, respectively. For the autoprotolysis of water, we used p$K_w = 14$.

|    | p$K_s'$    | p$K_0$    | p$K_1$    | p$K_2$   | p$K_{m0}$ |
|----|------------|-----------|-----------|----------|-----------|
| Fe | 4.9[58]    | 2.2[58]   | 3.5[58]   | 6.0[58]  | 8.6       |
| Al | 10.4[59]   | 5.0[59]   | 5.1[59]   | 6.6[59]  | 5.7       |
| Y  | 18.05[62]  | 7.7[60]   | 10.0      | 15.0     | 4.2       |

TABLE I. Values for the equilibrium constants and solubility products defined in the previous section. All values are given as decadic logarithm. Note that superscripts do not denote exponents, but specify the references. Values without reference have been estimated, since no precise literature value is given to our knowledge (see discussion in text).

### 1. Hydrolysis of metal ions

The used values are summarized in I. For the hydrolysis constants, systematic variations are expected due to temperature and total ionic strength[57]. The general trend and relation between the metal salts, however, is expected to be conserved. The values for p$K_2$ have additional uncertainties since the values will depend on the choice how to distinguish between $M(OH)_3$ and the corresponding insoluble solid state which might be still present in a cluster state[57].

For the case of $Y^{3+}$, the literature values for p$K_1$ and p$K_2$ have been estimated since no published values exist to our knowledge. However, the pH below 7 in the examined solutions makes these reactions negligible.

### 2. Solubility of metal hydroxides

Using the solubility product $K_s$, we obtain $K_s' = K_s/K_w^3 = [M^{3+}]_{max}/[H^+]^3$, which can be used to calculate the maximum concentration of free, unbound metal ions at a given pH: $[M^{3+}]_{max} = 10^{-pK_s'-3pH}$.

Solubilities vary by several orders of magnitude between solid phases with aging times over years (e.g., gibbsite $Al(OH)_3$ or different hydrous ferric oxides as hematite and geothite) and the amorphous state aged only for hours[58,59,61,62]. The experiments discussed in this study were performed within hours after preparation, suggesting the amorphous state after several hours as the suitable reference system. Accordingly, these solubility products $K'_s$ are listed in I.

During our experiments, no precipitation of metal hydroxide $M(OH)_3$ has been observed. Metal ions are introduced via a chloride salt. Most metal atoms occur in hydroxy complexes and thus lower the pH considerably. Furthermore, a considerable portion of the metal ions is removed from the solubility equilibrium due to association to the protein surface. During the calculation, the concentration of free, unbound ions was checked and found generally smaller than the corresponding maximum concentration $[M^{3+}]_{max}$. Thus, no metal hydroxide precipitation has to be taken into account.

### 3. Protonation and dissociation constants of amino acid functional groups

Protonation at the protein surface occurs mainly at the acidic amino acids, i.e. aspartic and glutamic acid. The related intrinsic equilibrium constants of the side chain in an isolated amino acid are $pK'_a = 3.71$ and $4.15$, respectively[63]. Protonation of basic amino acid residues arginine and lysine has $pK'_b = 12.10$ and $10.67$ (Ref.[63]). In both cases, however, it has to be kept in mind that solvation energies as well as the local electrostatic environment shift the equilibrium constants for individual sites considerably. To account for these variations is the central task of more detailed titration calculations[13,55].

Estimating shifts in $pK$ and aiming for representative binding constants for the pH range between 3 and 7, we use the following equilibrium constants: $pK_a = 5$ and $pK_b = 9$. The number of basic residues is set to 40. The number of acidic residues is set to 52, out of which 10 also can bind metal ions. The number of metal binding sites is chosen consistent with the semiempirical fit from the experimental section. The total numbers of functional groups are reasonable choices based on reported values for functional groups. BSA and HSA as well as dimers of BLG and OVA have approximately 80 basic and 90-100 acidic residues[39]. Note, though, that considerable part of these might not dissociate in the given pH range due to solvation shifts in the dissociation constant.

### 4. Association of metal ions to the protein

Several studies report equilibrium constants for the association of Y(III) to bone sialoprotein[64] as well as Fe(III) and Al(III) to transferrin[61,65,66]. These values have to be taken with care, since a single equilibrium constant subsumes all hydrolysis species of metal ions in one constant which thus presumably will be pH-dependent[61]. Furthermore, these studies mainly focussed on metal storage and transport proteins with specialized cavities. However, also solvent-exposed glutamates and aspartates have been shown to represent binding sites for multivalent cations with low affinity and low cation specificity[67].

Since no values for the association constants of metal ions to side chains are given in the literature, we used those reported for carboxylate groups of single amino acids in solution. The equilibrium constants vary with the actual amino acid and solution conditions like buffer salts as well as temperature. For $Fe^{3+}$, values for $pK_m$ range from 8.6 to 9.2; for $Al^{3+}$, values between 5.7 and 6.7 have been reported[68]. Even lower values are given for $Y^{3+}$, ranging between 4.4 and 5.4 (Ref.[68,69]).

These values do not include any solvation effects which are relevant for binding towards a carboxylic acid at the protein surface and in general shift all equilibrium constants to lower values. Furthermore, the values do not correspond directly to $pK_{m0}$, but incorporate binding of all metal hydroxy complexes simultaneously. $pK_{m1}$ and $pK_{m2}$ should be smaller due to a decreased electrostatic contribution to the binding. Since our approach is not aimed at describing the charging in full quantitative accuracy, but rather at qualitatively rationalizing the main effects, we choose the equilibrium constants $pK_{m0}$, $pK_{m1}$ and $pK_{m2}$ with the following estimation procedure: First, we choose $pK_{m0}$ as the lowest value from the published range as listed also in I. These values account already for the loss of solvation due to the binding site, i.e. one amino acid. Neighboring amino acids of the proteins will cause further loss of solvation; the effect, however, is not easy to estimate. Thus, in a second step, we decrease the $pK_{m0}$ by 2 to roughly account for the additional solvation effect in our course-grained model. Third, we estimate the pure electrostatic contribution to $pK_{m0}$ from the binding distance $r_b \approx 2$Å found in protein crystal structures[25]. This gives $\Delta pK_m \approx e^2/(4\pi\epsilon r_b k_B T \ln(10)) \approx 1.5$ per ligand charge. We thus obtain $pK_{m1}$ and $pK_{m2}$ by subtraction of 1.5 and 3 from $pK_{m0}$, respectively.

### C. Contributions to the free energy and surface charge effects

The equilibrium constants given in the previous sections can be rewritten as

$$K = \exp\left(\beta \Delta G\right) = \exp\left(\beta(\Delta G_{\text{intr}} + \Delta G_{\text{es}})\right) \quad (11)$$

Here we subsume all contributions under only two terms: first, the intrinsic term $\Delta G_{\text{intr}}$ accounts for short-range, non-coulombic contributions. In particular, this term includes localized contributions like hydrogen bonding and coordinative bonds, but also accounts for non-localized free energy contributions arising from hydrophobic binding, solvation effects and other water-mediated interactions[36]. Furthermore, entropic contributions and ion activities as well as free energy contributions due to ion correlations are included in the intrinsic part, since they show a comparable effect[37].

Second, $\Delta G_{\text{es}}$ accounts for the electrostatic long-range interaction of the global surface charge $Q$ with the ligand.

Using the surface potential $\Psi_0$, the terms reads

$$\Delta G_{es} = Ze\Psi_0 \quad (12)$$

where $Z$ is the formal charge of the ligand. In other words, the local density of ions varies according to the electrostatic potential at the reaction center and, thus, the association reaction will also depend on the protein charge. Eq. (4) defines the relation between the zeta potential $\Psi_0$ and the global surface charge $Q$. With this in mind, all binding constants have to be considered as depending on $Q$:

$$K_a = K'_a\,\tau(Q)^{-1}\,,\ K_b = K'_b\,\tau(Q)\,,\ K_{mi} = K'_{mi}\,\tau(Q)^{2-i}$$

where $K' = \exp(\beta\Delta G_{\text{intr}})$ denotes the intrinsic binding constant, respectively, and $\tau(Q) = \exp(\beta\,e\Psi_0(Q))$ represents the Boltzmann factor from the surface potential.

The surface charge $Q$ can be expressed in terms of the dissociated surface groups and bound metal ions (see Appendix, Eq. (13)). Since the amount of binding in turn depends on the association constants $K(Q)$ and thus the surface charge $Q$, rendering the full determination an implicit problem. The method to solve the system of equations applied for this work is explained in more detail in the Appendix.

## V. MODEL RESULTS

### A. Charge Inversion and solution's pH

5 summarizes the model results for a fixed protein concentration (BSA or HSA 2 mg/ml) with different salts. Both solution pH and protein charge vary faster for acidic salts - as intuitively expected and observed in the experiments (cf. 2, 3). In particular, for FeCl$_3$, the extent of charge inversion is much larger and the point of zero charge occurs at much lower salt concentration compared to YCl$_3$. While for YCl$_3$, the pH stays rather close to neutral within the experimental concentration range, it decreases below the isoelectric point, pI, for both other salts at rather low salt concentrations. Note that in both cases charge inversion occurs clearly before the pH value reaches pI.

When compared to the experiments, the qualitative agreement is very good. Even quantitatively, several quantities such as the point of zero charge and the salt concentration at the pI do not differ too much, indicating an appropriate choice for the model parameters.

### B. Phase Behavior: Reentrant Condensation from a DLVO Picture

The phase behavior can be estimated by calculating the stability from a DLVO-like picture[4,5] by using the surface charge as evaluated from the association model for the repulsive screened Coulomb interaction. We choose a Hamaker constant of $A = 3k_BT$ for the attractive van der Waals interaction[70]. The energy barrier is calculated as the difference between the maximum of the potential and the potential at equilibrium distance[71], $r_{\text{eq}} = 2\left(\frac{\pi}{\sqrt{18}}\frac{3}{4\pi n}\right)^{1/3} = \left(\frac{\sqrt{2}}{n}\right)^{1/3}$ with the protein number density $n$. If no maximum exists, the barrier is set to zero, representing phase regimes without stabilization against aggregation.

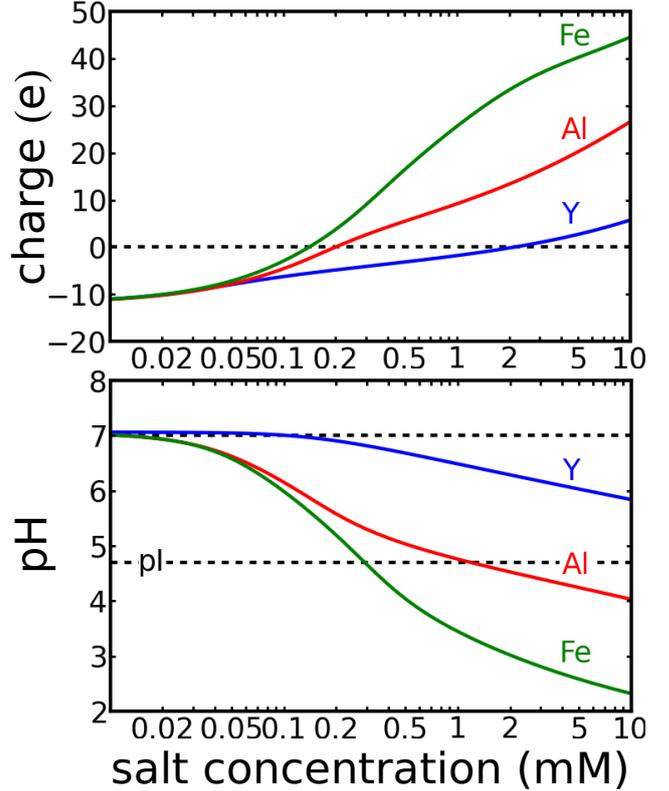

FIG. 5. Results from the binding model for BSA / HSA (2 mg/ml) with FeAl$_3$, AlCl$_3$ and YCl$_3$. Top: The inversion of protein surface charge occurs at much lower salt concentrations for acidic salts. Bottom: pH in general decrease upon addition of salt. While for YCl$_3$ the solutions stays nearly neutral, the other two salts cause the pH to decrease below the isoelectric point pI. Note that the model results show good agreement with the experimental results (2, 3).

6 presents the results in color coded form (in units of $k_BT$). In addition, profile lines for the surface charge are given (in units of elementary charges $e$). The general shape of the phase diagram (6) resembles obviously the experimental phase diagram (1). For all salts and protein concentrations, a charge inversion is found, which causes a reentrant charge stabilization and thus reentrant condensation. Furthermore, the phase diagram of acidic salts like AlCl$_3$ and FeCl$_3$ differs clearly from those found for YCl$_3$. As in the experimental findings, the phase-separated regime extends over a much larger salt concentration range for the case of YCl$_3$.

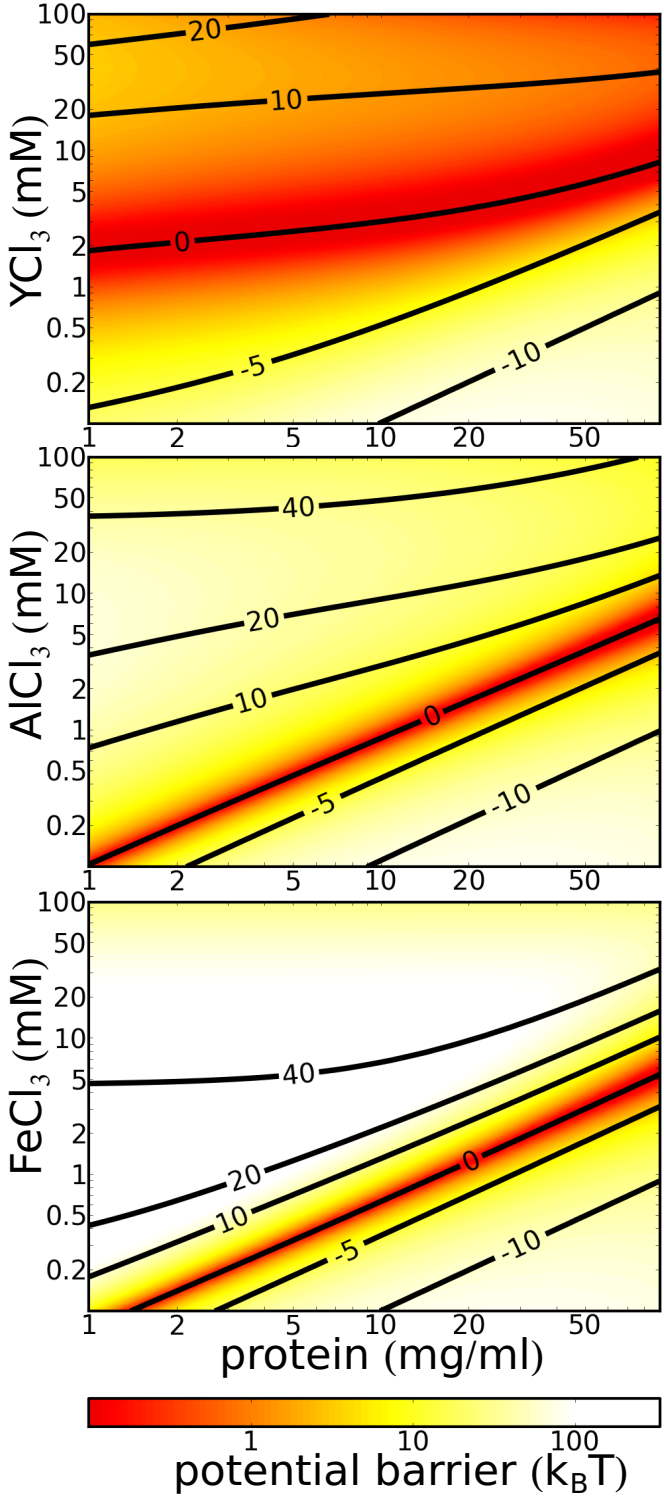

FIG. 6. Stability barriers (color-coded in units of $k_BT$) and protein charges (profile lines with indicated absolute charges in $e$) for the full parameter space of the different metal ions. In general, a reentrant phase behavior is observed with very similar topology as the experimental results (cf. 1). For acidic salts like $FeCl_3$, the charge inversion occurs at much lower salt concentration and within a smaller concentration range than for rather neutral salts like $YCl_3$.

## VI. DISCUSSION

### A. Mechanism of protein charge inversion

The experimental data for charge inversion have been explained via a physicochemical model. Interaction of multivalent metal ions with amino acid side chains occurs mostly in a very specific manner. Primarily responsible for this specificity is the metal ion chelation by carboxylate groups of Asp and Glu. There is clear evidence for this interaction mechanism in one of the systems studied here. A recently published[72] crystal structure of BLG in the presence of $Y^{3+}$ reveals four well-defined binding sites for $Y^{3+}$ cations. The cations coordinate in multidentate geometries with up to four carboxylate groups, thereby acting as cross-linkers between the protein molecules.

The reentrant condensation of proteins could be qualitatively described using Monte-Carlo simulations for the binding of cations to acidic amino acids exposed to the solvent[38,39]. A recent study investigated the molecular mechanism of charge inversion in solutions of tetra-aspartate, representing a model "protein" in the sense of exposing carboxylic acids to the solvent[73]. Mono- and bidentate association geometries of cations at carboxylate groups were found, indicating ion-carboxyl pairing to be the underlying mechanism. Ion-protein interaction was found to be dominated by electrostatic attraction between cations and the carboxyl group, while ion specific chemical contributions seemed less important[73].

Protonation equilibria of charged amino acid side chains are the second factor contributing to charge inversion. Variations of the pH cause considerable contributions to the surface charge. This effect is particularly pronounced in the case of metal ions like $Fe^{3+}$ and $Al^{3+}$, which cause a significant pH shift by hydrolysis themselves. Although a generic effect of proteins, the extent is dependent on the specific protein and its structure, in particular the number of acidic and basic residues. For unbuffered solutions, the specific combination of protein and salt thus can have an important effect on the results. A recent simulation study based on a course-grained protein model found a dependence of the isoelectric point on monovalent ion concentration[16], supporting the notion that salt binding in combination with charge regulation governs the protein surface charge.

### B. Relation of charge inversion to reentrant condensation

Reentrant condensation has been suggested to be a universal phenomenon in solutions of several globular proteins with negative native charge in the presence of multivalent cations[39]. From a naïve point of view, charge inversion represents the essential mechanism allowing residual attraction to cause phase separation, once the Coulombic repulsion vanishes close to zero protein net charge.

Recent results inspire a more refined relation of reentrant condensation and charge inversion.

For BLG with divalent salts the solutions do not clear up completely for high salt concentrations, although a significant charge inversion has occured. Similarly, reentrant condensation is found to be incompletely established for solutions of glucose isomerase in the presence of $Y^{3+}$, whereas a charge inversion is present. Combining these finding with the presented results of this study, charge inversion is found to be a universal phenomenon for globular proteins with negative native charge in the presence of multivalent cations, while the occurence of the reentrant regime seems to depend on other more specific conditions.

This finding is also in accordance with the observation that charge inversion caused by titration does not induce reentrant condensation. For both HSA and BSA, solutions remain stable when reducing the pH way below the isoelectric point, even when passing a conformational transition (normal to fast form)[52,74,75].

The protein charge controls the Coulombic repulsion between the protein molecules. For the case of salt-induced charge inversion, the molecules repel each other at low and high salt concentrations, whereas no Coulombic repulsion is present at the point of zero charge. The difference in phase behavior, i.e. the occurence of the reentrant regime, mainly arises from the strength and nature of attraction in the system. Since several proteins without multivalent ions do not phase separate at the isoelectric point, the binding of multivalent cations does not only cause charge inversion, but also causes additional attraction between the proteins. Possible mechanism for this attraction could be cation-crosslinks between two protein molecules[72] or attractive patches due to charge and hydrophobicity patterns.

These findings also imply another hypothetical mechanism besides the pH effect for the narrowing of the condensation regime for $AlCl_3$ and $FeCl_3$ compared to $YCl_3$. For acidic salts, the variation of pH causes a significant part of the charge inversion due to side chain protonation. The number of bound cations at the point of zero overall charge is decreased, implying that ion-induced attractions like cross-linking or patch formation are less established. Less attraction in turn means a narrower range around the point of zero charge, where repulsion is small enough to allow sensitive phase separation and aggregation. Thus, variation of the pH might not only influence the charge regulation, but could also be a control parameter to further fine tune the protein interactions, both attractive and repulsive, in order to allow control of phase behavior and nucleation conditions[41].

## VII. CONCLUSION

Using electrophoretic light scattering, charge inversion of globular proteins with negative native charge in the presence of multivalent metal cations has been shown to be a common phenomenon. Effects of the pH arising from the hydrolysis of different metal salts induce significant changes for charge inversion and reentrant phase behavior. An analytical model based on chemical association reactions and metal hydrolysis provides good qualitative and even reasonable quantitative agreement with the experimental data, supporting the view of general features governing the charge inversion in protein solutions. Both local pairing of ions and carboxyl groups and protonation represent likely mechanisms dominating the charge inversion. Thus, in contrast to charge inversion in other biological molecules such as DNA, protein charge inversion should mainly be considered as a phenomenon driven by ion-protein pairing. The actual extent of charge inversion is determined by the specific combination of the protein and added salt, rendering a detailed prediction rather difficult.

For the phenomenon of reentrant condensation, the picture becomes even more involved. Multivalent cations do not only compensate the protein charge, but also provide attractive interaction patches between proteins. These attractive sites can be represented by crosslinks of multivalent cations as found in protein crystals as well as by varied surface patterns of charges and hydrophobicity. The interplay of cation binding and pH, thus, not only governs the protein surface charge, but also governs at least part of the protein interactions. The control of pH in combination with multivalent cations is suggested as a methodological improvement to control phase and nucleation behavior of protein solutions. In a broader context, the interplay of pH and ion association is relevant for a comprehensive understanding of biological and soft matter systems in aqueous electrolyte solutions.

## APPENDIX: MODEL FOR ION ASSOCIATION AND pH EFFECTS: CALCULATION METHOD

In this appendix we present the model combining protein-ion interactions and protolysis equilibria. Starting from conservation laws for the number densities of all involved components, we transform the system of nonlinear equation to a polynomial in $[H^+]$, which has been found to have only one physical solution. The charge effect is implemented by a numeric bisection approach with respect to the surface charge.

## A. Conservation Laws

The reactions in Eqn. (7–10) imply conservation laws for the total number density of all components:

$$m = \sum_n [\text{M(OH)}_n^{3-n}] + \sum_n [\text{M(OH)}_n \text{A}^{2-n}]$$
$$a = [\text{A}^-] + [\text{HA}] + \sum_n [\text{M(OH)}_n \text{A}^{2-n}]$$
$$d = [\text{D}^-] + [\text{HD}]$$
$$b = [\text{B}^+] + [\text{B(OH)}] \ .$$

For simplicity of notation, we use $m$ for the total metal ion concentration (bound and free), $b$ for the total basic residue concentration and $a$ and $d$ for the acidic redidue concentrations (ion-binding and inert). The charge neutrality of the system implies a fifth equation incorporating all charged species:

$$0 = \sum_n \left((3-n)[\text{M(OH)}_n^{3-n}] + (2-n)[\text{M(OH)}_n \text{A}^{2-n}]\right) \cdots$$
$$-3m + a - [\text{A}^-] + [\text{B}^+] - b + d - [\text{D}^-] + [\text{H}^+] - [\text{OH}^-] \ ,$$

including monovalent ions from the dissociation of the salt ($3m$) and the amino acids of the protein ($a, d, b$), which are assumed not to have a significant influence on the initial pH at zero salt concentration.

From mass action we obtain the following polynomials in $[\text{H}^+]$, which summarize the total concentrations of different species occuring in the reactions Eqn. (7–10):

$$\alpha([\text{H}^+], K_i) = \sum_{j=0}^{3} [\text{H}^+]^j \prod_{n=3-j}^{2} K_n$$
$$\beta([\text{H}^+], K_i, L_i) = \sum_{j=0}^{3} L_{3-j} [\text{H}^+]^j \prod_{n=3-j}^{2} K_n$$
$$\gamma([\text{H}^+], K_a) = 1 + [\text{H}^+] K_a$$
$$\delta([\text{H}^+], K_b) = 1 + [\text{H}^+] K_b$$
$$\epsilon([\text{H}^+], K_i, L_i) = \sum_{j=0}^{3} (j-1) L_{3-j} [\text{H}^+]^j \prod_{n=3-j}^{2} K_n$$
$$\phi([\text{H}^+], K_i) = \sum_{j=0}^{3} j \, [\text{H}^+]^j \prod_{n=3-j}^{2} K_n$$
$$\eta(v[\text{H}^+], K_0) = (K_0 [\text{H}^+])$$

Using these polynomials, the conservation laws can be rewritten in matrix form and the equation system can be transformed into a set of four linear and one non-linear equations:

$$\begin{pmatrix} m \\ a \\ d \\ b \\ h' \end{pmatrix} = \begin{pmatrix} \alpha & 0 & 0 & 0 & \beta \\ 0 & \gamma & 0 & 0 & \beta \\ 0 & 0 & \gamma & 0 & 0 \\ 0 & 0 & 0 & \delta & 0 \\ -\eta\phi & \eta & \eta & (1-\delta)\eta & -\epsilon\eta \end{pmatrix} \cdot \begin{pmatrix} [\text{M(OH)}_3] \\ [\text{A}^-] \\ [\text{D}^-] \\ [\text{B(OH)}] \\ [\text{A}^-][\text{M(OH)}_3] \end{pmatrix} ,$$

$$\chi \begin{pmatrix} [\text{M(OH)}_3] \\ [\text{A}^-] \\ [\text{D}^-] \\ [\text{B(OH)}] \\ [\text{A}^-][\text{M(OH)}_3] \end{pmatrix} =$$

$$\begin{pmatrix} \gamma\delta\eta(\beta+\epsilon\gamma) & -\beta\gamma\delta\eta & -\beta\gamma\delta\eta & -\beta(1-\delta)\eta\gamma^2 & \beta\delta\gamma^2 \\ \beta\gamma\delta\eta\phi & \gamma\delta\eta(\alpha\epsilon-\beta\phi) & -\alpha\beta\delta\eta & -\alpha\beta\gamma(1-\delta)\eta & \alpha\beta\gamma\delta \\ 0 & 0 & \delta\eta(\alpha(\beta+\epsilon\gamma)-\beta\gamma\phi) & 0 & 0 \\ 0 & 0 & 0 & \gamma\eta(\alpha(\beta+\epsilon\gamma)-\beta\gamma\phi) & 0 \\ -\delta\eta\gamma^2\phi & \alpha\gamma\delta\eta & \alpha\delta\eta\gamma & \alpha(1-\delta)\eta\gamma^2 & -\alpha\delta\gamma^2 \end{pmatrix} \cdot \begin{pmatrix} m \\ a \\ d \\ b \\ h' \end{pmatrix}$$

where we use the short-hand notations $h' := -1 + ([\text{H}^+] - 3m + (a+d) - b)\eta$ and $\chi = \gamma\delta\eta(\alpha\beta + \alpha\gamma\epsilon - \beta\gamma\phi)$. We used *Wolfram Mathematica® 8* to perform this task.

## B. Derivation of polynomial in $[\text{H}^+]$ and physical solution

Equating the product of the first and second row with the fifth row, we obtain a polynomial in $[\text{H}^+]$:

$$0 = \left(\chi [\text{A}^-]\right)_{\text{row2}} \cdot \left(\chi [\text{M(OH)}_3]\right)_{\text{row1}}$$
$$-\chi \left(\chi [\text{A}^-][\text{M(OH)}_3]\right)_{\text{row5}}$$
$$=: f_{(m,a,b,d)} \left([\text{H}^+]\right) \ .$$

Any solution of the system of equations has to be a root of $f_{(m,a,b,d)}([H^+])$. Given a value of $[H^+]$, all other concentrations of reaction species can be calculated. The physical solution is obtained by testing straight-forward physical conditions for the mathematical roots $[H^+]$, returning under all studied conditions a unique solution:

(i) $[H^+] \in \mathbb{R}^+$

(ii) $m > [M(OH)_3]_{[H^+],m,a,b,d} > 0$

(iii) $a > [A^-]_{[H^+],m,a,b,d} > 0$

(iv) $b > [B(OH)]_{[H^+],m,a,b,d} > 0$

(v) $d > [D^-]_{[H^+],m,a,b,d} > 0$

### C. Surface Charge Effect on Association and Implicit Relation

As pointed out in the main text, surface charge influences the association constants and is in turn determined by the amount of association. Using the solution for $[H^+]$ from the previous subsection, the surface charge can be calculated for a given set of association constants $K_a, K_b, K_w, K_i, L_i$:

$$Q = [B^+] - [A^-] - [D^-] + \sum_n (2-n) \cdot [M(OH)_n A^{2-n}]$$
$$=: g([H^+], m, a, b, d)$$

In order to solve this implicit problem, we performed a numerical bisection with respect to $Q$ for the function $g([H^+], m, a, b, d)\big|_Q - Q$ in the interval $\left[0, g([H^+], m, a, b, d)\big|_{Q=0}\right]$ using the routine supplied by the Python package *scipy.optimize*[50,51] (version 0.7.0).

### ACKNOWLEDGEMENTS

The authors acknowledge financial support by the Deutsche Forschungsgemeinschaft (DFG). FRR acknowledges a fellowship by the Studienstiftung des Deutschen Volkes. The authors benefitted from discussions with M. Hennig (ILL, Grenoble, France), O. Lenz, C. Holm (University of Stuttgart, Germany), M. Antalik (Slovak Academy of Sciences, Kosice), and P. Jungwirth (Academy of Sciences of the Czech Republic, Prague).